\documentclass[aps,pra,groupedaddress,twocolumn]{revtex4-1}
\usepackage{amsmath}
\usepackage{amssymb}
\usepackage{graphicx}	
\usepackage{bm} 
\usepackage{color}
\usepackage{epstopdf}

\begin{document}
\title{Quantum few-body bound states of dipolar particles in a helical geometry}

\author{J.~K. Pedersen}
\author{D.~V. Fedorov}
\author{A.~S. Jensen}
\author{N.~T. Zinner}
\affiliation{Department of Physics and Astronomy, Aarhus University, DK-8000 Aarhus C, Denmark}

\date{\today}

\begin{abstract}
We study a quantum mechanical system consisting of up to three identical dipoles confined to move along a helical shaped trap. 
The long-range interactions between particles confined to move in this one dimension leads to an interesting effective two-particle potential with an oscillating behaviour. 
For this system we calculate the spectrum and the wave functions of the bound states. The full quantum solutions show clear imprints of the tendency for the 
system to form chains of dipoles along the helix, i.e. a configuration in which the dipoles are sitting approximately one winding of the helix apart so that they can take maximal advantage of the strong head-to-tail attraction that is a generic feature of the dipole-dipole interaction.
\end{abstract}

\maketitle

\section{Introduction}
Cold atomic gas physics has advanced to the stage at which simulation and exploration of quantum phenomena can now be 
done at unprecedented levels of accuracy and with great potential for engineering systems that are otherwise very 
hard to access in other fields of research \cite{lewenstein2007,bloch2008,esslinger2010,zinner2013,zinner2014}.
An important recent advance is the ability to create homo- and heteronuclear molecules or trap atoms with large permanent magnetic moments at low temperatures which has many interesting applications \cite{lahaye2009,carr2009,baranov2012,yan2013,barry2014,aikawa2014,frisch2014,maier2015,paz2015}. 
The head-to-tail attraction between such molecules with dipole moments is a concern, as it limits the timescales at which an experiment can operate due to strong loses. One way to solve this is to confine the molecules in lower dimensional traps, where the head-to-tail attraction can be suppressed \cite{miranda2011,chotia2012}. 

The long-range dipole-dipole interaction has a very interesting interplay with the geometry of the lower dimensional systems. Here we consider the one dimensional geometry of a helix. This kind of system can be realized with Laguerre-Gaussian beams with nonzero angular momentum \cite{macdonald2002,pang2005,bhatta2007,ricardez2010,okulov2012,arnold2012,beattie2013} or by trapping the atoms in the evanescent field surrounding an optical fiber\cite{sague2008,vetsch2010,dawkins2011,reitz2012}. The detailed interplay between
the non-trivial long-range interaction of both charged ions and dipolar particles on a helical geometry has given rise to a number
of theoretical studies into these setups 
\cite{law2008,huhta2010,schmelcher2011,zampetaki2013,pedersen2014,stockhofe2014,zampetaki2015a,stockhofe2015,zampetaki2015b}.

In this work we consider dipolar particles on a single helix which confines the movement of the dipoles. We will 
be assuming that the helical trap is strongly confined in the directions that are locally perpendicular to the 
helix such that we may ignore the motion of the particles in these transverse directions. In this setup we investigate
the formation of quantum mechanical bound states of two and three particles on the helix, we study their structure and 
the criteria for formation. The paper is organized as follows. In section II we introduce the helix and the dipole-dipole interaction. In section III we look at two dipoles on a helix, we calculate the bound states and look at their energy and their size. In section IV we introduce another dipole and look at the bound states of three dipoles on a helix. We conclude in section V. 

\section{The Helix Model}
\begin{figure}
\includegraphics[scale=1.0]{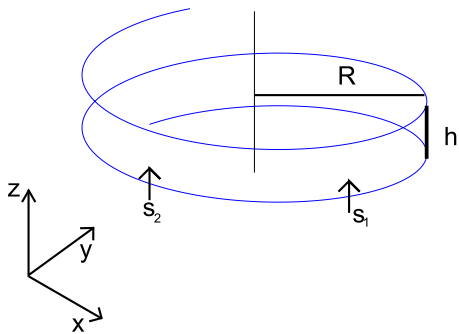}
\caption{A representation of a helix of radius $R$ and pitch $h$,
  with dipoles positioned at $s_1$ and $s_2$, respectively. Here $s$ is the arc length
	measured along the helix.
	We assume that an external field 
	is applied which directs the dipole moments along the
  $z$-axis. The cartesian axes are indicated on the left at the bottom. \label{helix}}
\end{figure}

A helix is defined by two parameters, the radius $R$, and the pitch $h$. The radius defines the circumference while the 
pitch defines how far one moves along the symmetry axis of the helix during one revolution. The geometry of the helical 
setup is illutrated in figure~\ref{helix}.
To describe the position along the helix it is convenient to use the arc length $s$ measured along the helix, but for numerical purposes we will instead use the angle around the central symmetry axis $\phi=s/\alpha$, where $\alpha=\sqrt{R^2+\left(\frac{h}{2\pi}\right)^2}$. It is worth noting, that this angle $\phi$ is not limited to be smaller than $2\pi$, anything more than that just means more than one revolution on the helix.  
The relation between the coordinates on the helix, and the Cartesian coordinates in three dimensional space is
\begin{equation}
 (x,y,z) = (R\sin \phi,R\cos \phi,h \frac{\phi}{2\pi}),
\label{eq:helixeq}
\end{equation}
where $R$ is positive and $\phi$ runs through an interval between $\phi_{min}$ and $\phi_{max}$. On the helix we place between one and three dipoles, all of equal mass $m$, and dipole moment $\mathbf{d}$, all the dipoles are aligned by an external field. 

The long-range interaction of two particles on the helix are given by the usual dipole-dipole interaction 
between two dipoles in three-dimensional space.
The two-body interaction potential, $V$, between two dipoles at position $\bm{r}_i$ and
$\bm{r}_j$ is
\begin{equation}
 V(\bm{r}_i,\bm{r}_j) = \frac{1}{4\pi\epsilon_0}\frac{1}{r^3}\left[\bm{d}
 \cdot\bm{d}-3\left(\bm{d}\cdot\bm{\hat{r}}\right)\left(\bm{d}
 \cdot\bm{\hat{r}}\right)\right],
\label{eq:3dpotential}
\end{equation}
where $r=|\bm{r}_i - \bm{r}_j|$ is the distance between the dipoles,
and $\bm{\hat{r}} = (\bm{r}_i - \bm{r}_j)/r $ is the unit vector in the
direction connecting the two dipoles. We use the unit $\frac{1}{4\pi\epsilon_0}$, where $\epsilon_0$ is the permitivity, corresponding to electric dipoles, but in principle the formalism is applicable for magnetic dipoles if we use the unit $\frac{\mu_0}{4\hbar}$, where $\mu_0$ is the vacuum permeability. Here we will assume that an external field is used to align the 
dipole moments of all particles along the symmetry axis which we take to be the $z$-axis (see figure~\ref{helix}).
As mentioned above, instead of describing the position of the dipoles by their Cartesian coordinates it turns out to
be easier to use their position along the helix. This can be done 
through the coordinate transformation in Eq.~\eqref{eq:helixeq}. 
If the position of the two dipoles along the helix are $\phi_i$ and $\phi_j$ 
the two-dipole potential in Eq.~\eqref{eq:3dpotential} becomes
\begin{eqnarray}
 & & V(\phi_i,\phi_j)=\frac{d^2}{4\pi\epsilon_0}  \label{eq:potential} 
 \\ \nonumber
  &\times& \frac{2R^2\left[1-
 \cos{(\phi_i-\phi_j)}\right]
 -2 h^2\left((\phi_i-\phi_j)/(2\pi) \right)^2 }
 {\left( 2R^2\left[1-\cos{(\phi_i-\phi_j)}\right]
 + h^2\left((\phi_i-\phi_j)/(2\pi) \right)^2 
 \right)^{5/2}}.
\end{eqnarray} 
We see that for dipoles trapped on the helix with dipole moments pointing in the the $z$-direction the two-dipole potential 
in Eq.~\eqref{eq:potential} only depends on the distance between the two dipoles $\phi=\phi_i-\phi_j$ measured along the helix. 

\begin{figure}
\centering
\begingroup
  \makeatletter
  \providecommand\color[2][]{%
    \GenericError{(gnuplot) \space\space\space\@spaces}{%
      Package color not loaded in conjunction with
      terminal option `colourtext'%
    }{See the gnuplot documentation for explanation.%
    }{Either use 'blacktext' in gnuplot or load the package
      color.sty in LaTeX.}%
    \renewcommand\color[2][]{}%
  }%
  \providecommand\includegraphics[2][]{%
    \GenericError{(gnuplot) \space\space\space\@spaces}{%
      Package graphicx or graphics not loaded%
    }{See the gnuplot documentation for explanation.%
    }{The gnuplot epslatex terminal needs graphicx.sty or graphics.sty.}%
    \renewcommand\includegraphics[2][]{}%
  }%
  \providecommand\rotatebox[2]{#2}%
  \@ifundefined{ifGPcolor}{%
    \newif\ifGPcolor
    \GPcolortrue
  }{}%
  \@ifundefined{ifGPblacktext}{%
    \newif\ifGPblacktext
    \GPblacktextfalse
  }{}%
  \let\gplgaddtomacro\g@addto@macro
  \gdef\gplbacktext{}%
  \gdef\gplfronttext{}%
  \makeatother
  \ifGPblacktext
    \def\colorrgb#1{}%
    \def\colorgray#1{}%
  \else
    \ifGPcolor
      \def\colorrgb#1{\color[rgb]{#1}}%
      \def\colorgray#1{\color[gray]{#1}}%
      \expandafter\def\csname LTw\endcsname{\color{white}}%
      \expandafter\def\csname LTb\endcsname{\color{black}}%
      \expandafter\def\csname LTa\endcsname{\color{black}}%
      \expandafter\def\csname LT0\endcsname{\color[rgb]{1,0,0}}%
      \expandafter\def\csname LT1\endcsname{\color[rgb]{0,1,0}}%
      \expandafter\def\csname LT2\endcsname{\color[rgb]{0,0,1}}%
      \expandafter\def\csname LT3\endcsname{\color[rgb]{1,0,1}}%
      \expandafter\def\csname LT4\endcsname{\color[rgb]{0,1,1}}%
      \expandafter\def\csname LT5\endcsname{\color[rgb]{1,1,0}}%
      \expandafter\def\csname LT6\endcsname{\color[rgb]{0,0,0}}%
      \expandafter\def\csname LT7\endcsname{\color[rgb]{1,0.3,0}}%
      \expandafter\def\csname LT8\endcsname{\color[rgb]{0.5,0.5,0.5}}%
    \else
      \def\colorrgb#1{\color{black}}%
      \def\colorgray#1{\color[gray]{#1}}%
      \expandafter\def\csname LTw\endcsname{\color{white}}%
      \expandafter\def\csname LTb\endcsname{\color{black}}%
      \expandafter\def\csname LTa\endcsname{\color{black}}%
      \expandafter\def\csname LT0\endcsname{\color{black}}%
      \expandafter\def\csname LT1\endcsname{\color{black}}%
      \expandafter\def\csname LT2\endcsname{\color{black}}%
      \expandafter\def\csname LT3\endcsname{\color{black}}%
      \expandafter\def\csname LT4\endcsname{\color{black}}%
      \expandafter\def\csname LT5\endcsname{\color{black}}%
      \expandafter\def\csname LT6\endcsname{\color{black}}%
      \expandafter\def\csname LT7\endcsname{\color{black}}%
      \expandafter\def\csname LT8\endcsname{\color{black}}%
    \fi
  \fi
  \setlength{\unitlength}{0.0500bp}%
  \begin{picture}(4676.00,4534.00)%
    \gplgaddtomacro\gplbacktext{%
      \csname LTb\endcsname%
      \put(946,704){\makebox(0,0)[r]{\strut{}-2.5}}%
      \put(946,1213){\makebox(0,0)[r]{\strut{}-2.0}}%
      \put(946,1723){\makebox(0,0)[r]{\strut{}-1.5}}%
      \put(946,2232){\makebox(0,0)[r]{\strut{}-1.0}}%
      \put(946,2741){\makebox(0,0)[r]{\strut{}-0.5}}%
      \put(946,3250){\makebox(0,0)[r]{\strut{} 0.0}}%
      \put(946,3760){\makebox(0,0)[r]{\strut{} 0.5}}%
      \put(946,4269){\makebox(0,0)[r]{\strut{} 1.0}}%
      \put(1538,484){\makebox(0,0){\strut{} 0.5}}%
      \put(2049,484){\makebox(0,0){\strut{} 1.0}}%
      \put(2559,484){\makebox(0,0){\strut{} 1.5}}%
      \put(3070,484){\makebox(0,0){\strut{} 2.0}}%
      \put(3581,484){\makebox(0,0){\strut{} 2.5}}%
      \put(4092,484){\makebox(0,0){\strut{} 3.0}}%
      \put(176,2486){\rotatebox{-270}{\makebox(0,0){\strut{}$\Tilde{V}(\phi)$}}}%
      \put(2678,154){\makebox(0,0){\strut{}$\frac{\phi}{2\pi}$}}%
    }%
    \gplgaddtomacro\gplfronttext{%
    }%
    \gplbacktext
    \put(0,0){\includegraphics{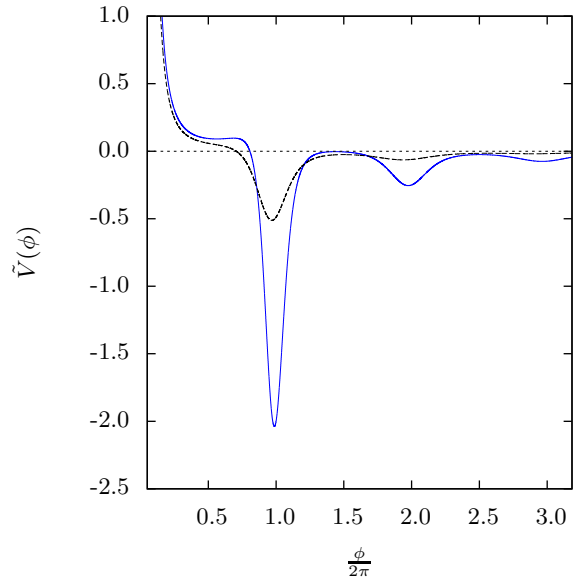}}%
    \gplfronttext
  \end{picture}%
\endgroup\caption{The reduced potential, $\tilde{V}=4\pi R^3 \epsilon_0 V /d^2$, of two
  dipoles as a function of the relative angle, $\phi =
  \phi_i-\phi_j$, separating the particles on the helix. The helix
  parameters are chosen to be $h = R$ (blue solid) and $h = 1.6 R$ (black dashed).  \label{fig:potential}}
\end{figure}

The two-dipole potential as a function of this dipole separation, is shown in figure~\ref{fig:potential}. The long range part of the potential falls of as $\phi^{-3}$ like the distance dependence from Eq.~\eqref{eq:3dpotential} in three dimensions. But the potential also contains oscillations at smaller distances. They disappear when $\frac{\phi}{2\pi}>1+(2\pi)^2\frac{R}{h}$. The corresponding minima in the potential arise whenever the dipoles are above each other. It is worth noting that the minima are not exactly at multipla of $2\pi$, which would correspond to the dipoles being directly above each other. Instead the minima are located at a distance from each other that is slightly smaller than $2\pi$ as it is a compromise between optimizing the angle, and minimizing the three dimensional distance between the dipoles.

When two dipolar particles come very close to each other we expect that strong interactions will cause for instance chemical reactions and losses from our setup. This is not the regime we are interested in and we therefore look for a parameter regime in which the short-range behavior of the dipole-dipole potential is purely repulsive. In this regime, we may safely ignore any chemical reactions and strong losses. This requirement puts limitations on the geometry of the helix. If $\frac{h}{R}>\sqrt{2}\pi$, then the two-dipole potential becomes attractive, and would allow distances where f.x. van der Waals forces would dominate. We thus restrict ourselves to cases where $\frac{h}{R}<\sqrt{2}\pi$.

\section{The two-body problem}
We now place two dipoles on the helix. We assume that they are only subject to their mutual interaction which only depend on the relative distance between the dipoles as in Eq.(\ref{eq:potential}).
It is convenient to make a change of variables, from the positions of the two dipoles $\phi_1$ and $\phi_2$, to relative angular distance $\phi=\phi_1-\phi_2$, and center of mass $\Phi=\phi_1+\phi_2$. 
A straightforward if somewhat tedious calculation now shows that the 
relative and center of mass motion decouples in the Hamiltonian. 
The center of mass motion will be governed by a free particle 
Hamiltonian which can be trivially solved for given boundary
conditions. In this paper we will ignore the center of mass
motion which is not important for the question of bound state formation.
The Hamiltonian governing the relative motion of the two dipoles takes the form
\begin{equation}
H=\frac{-\hbar^2}{2\mu\alpha^2}\frac{\partial^2}{\partial\phi^2}+V(\phi)
\end{equation}
Here $\mu=m/2$ is the reduced mass of the two dipoles. Note that one must be careful 
with quantization in geometries with non-trivial curvature. However, the regular helix 
we consider in this paper has the property that its curvature is in fact constant and
one may thus do the transformation from Cartesian to curvilinear coordinates in the 
standard manner \cite{stockhofe2014,knorborg2015}.
It is convenient to write the Hamiltonian in natural units, $\tilde H$, which has the form
\begin{equation}\label{twoH}
\tilde{H}=\frac{\mu\alpha^2}{\hbar^2}H=\frac{-1}{2}\frac{\partial^2}{\partial\phi^2}+\beta\tilde{V}(\phi)
\end{equation}
where $\beta$ is the potential strength given by
\begin{equation}
\beta=\frac{\mu d^2}{2\pi\epsilon_0R\hbar^2}\left(\frac{\alpha}{R}\right)^2
\end{equation}
and $\tilde{V}$ is the reduced potential where
\begin{equation}
\tilde{V}(\phi)=\frac{1-cos{\phi}-\left(\frac{h\phi}{2\pi R}\right)^2}{\left(2\left[1-\cos{\phi}\right]+\left(\frac{h\phi}{2\pi R}\right)^2\right)^{5/2}}.
\label{eq:Vphi}
\end{equation}
We thus measure lengths in units of $\alpha$ and energies in units of $\hbar^2/\mu\alpha^2$.
Because $V(\phi)\rightarrow\infty$ as $\phi\rightarrow 0$, the wavefunction $\psi(\phi)$ has to be zero at $\phi=0$. This means, that we only need to calculate the wavefunction for $\phi>0$, and then construct the wavefunction to be either odd or even depending on whether dipoles are identical fermions or bosons. 

\subsection{Solutions to the two-dipole system}
\begin{figure}
\centering
\begingroup
  \makeatletter
  \providecommand\color[2][]{%
    \GenericError{(gnuplot) \space\space\space\@spaces}{%
      Package color not loaded in conjunction with
      terminal option `colourtext'%
    }{See the gnuplot documentation for explanation.%
    }{Either use 'blacktext' in gnuplot or load the package
      color.sty in LaTeX.}%
    \renewcommand\color[2][]{}%
  }%
  \providecommand\includegraphics[2][]{%
    \GenericError{(gnuplot) \space\space\space\@spaces}{%
      Package graphicx or graphics not loaded%
    }{See the gnuplot documentation for explanation.%
    }{The gnuplot epslatex terminal needs graphicx.sty or graphics.sty.}%
    \renewcommand\includegraphics[2][]{}%
  }%
  \providecommand\rotatebox[2]{#2}%
  \@ifundefined{ifGPcolor}{%
    \newif\ifGPcolor
    \GPcolortrue
  }{}%
  \@ifundefined{ifGPblacktext}{%
    \newif\ifGPblacktext
    \GPblacktextfalse
  }{}%
  \let\gplgaddtomacro\g@addto@macro
  \gdef\gplbacktext{}%
  \gdef\gplfronttext{}%
  \makeatother
  \ifGPblacktext
    \def\colorrgb#1{}%
    \def\colorgray#1{}%
  \else
    \ifGPcolor
      \def\colorrgb#1{\color[rgb]{#1}}%
      \def\colorgray#1{\color[gray]{#1}}%
      \expandafter\def\csname LTw\endcsname{\color{white}}%
      \expandafter\def\csname LTb\endcsname{\color{black}}%
      \expandafter\def\csname LTa\endcsname{\color{black}}%
      \expandafter\def\csname LT0\endcsname{\color[rgb]{1,0,0}}%
      \expandafter\def\csname LT1\endcsname{\color[rgb]{0,1,0}}%
      \expandafter\def\csname LT2\endcsname{\color[rgb]{0,0,1}}%
      \expandafter\def\csname LT3\endcsname{\color[rgb]{1,0,1}}%
      \expandafter\def\csname LT4\endcsname{\color[rgb]{0,1,1}}%
      \expandafter\def\csname LT5\endcsname{\color[rgb]{1,1,0}}%
      \expandafter\def\csname LT6\endcsname{\color[rgb]{0,0,0}}%
      \expandafter\def\csname LT7\endcsname{\color[rgb]{1,0.3,0}}%
      \expandafter\def\csname LT8\endcsname{\color[rgb]{0.5,0.5,0.5}}%
    \else
      \def\colorrgb#1{\color{black}}%
      \def\colorgray#1{\color[gray]{#1}}%
      \expandafter\def\csname LTw\endcsname{\color{white}}%
      \expandafter\def\csname LTb\endcsname{\color{black}}%
      \expandafter\def\csname LTa\endcsname{\color{black}}%
      \expandafter\def\csname LT0\endcsname{\color{black}}%
      \expandafter\def\csname LT1\endcsname{\color{black}}%
      \expandafter\def\csname LT2\endcsname{\color{black}}%
      \expandafter\def\csname LT3\endcsname{\color{black}}%
      \expandafter\def\csname LT4\endcsname{\color{black}}%
      \expandafter\def\csname LT5\endcsname{\color{black}}%
      \expandafter\def\csname LT6\endcsname{\color{black}}%
      \expandafter\def\csname LT7\endcsname{\color{black}}%
      \expandafter\def\csname LT8\endcsname{\color{black}}%
    \fi
  \fi
  \setlength{\unitlength}{0.0500bp}%
  \begin{picture}(5102.00,3968.00)%
    \gplgaddtomacro\gplbacktext{%
      \csname LTb\endcsname%
      \put(946,704){\makebox(0,0)[r]{\strut{}-0.2}}%
      \put(946,1304){\makebox(0,0)[r]{\strut{} 0}}%
      \put(946,1904){\makebox(0,0)[r]{\strut{} 0.2}}%
      \put(946,2503){\makebox(0,0)[r]{\strut{} 0.4}}%
      \put(946,3103){\makebox(0,0)[r]{\strut{} 0.6}}%
      \put(946,3703){\makebox(0,0)[r]{\strut{} 0.8}}%
      \put(1078,484){\makebox(0,0){\strut{} 0}}%
      \put(1803,484){\makebox(0,0){\strut{} 20}}%
      \put(2529,484){\makebox(0,0){\strut{} 40}}%
      \put(3254,484){\makebox(0,0){\strut{} 60}}%
      \put(3980,484){\makebox(0,0){\strut{} 80}}%
      \put(4705,484){\makebox(0,0){\strut{} 100}}%
      \put(176,2203){\rotatebox{-270}{\makebox(0,0){\strut{}$\psi(\phi)$}}}%
      \put(2891,154){\makebox(0,0){\strut{}$\phi$}}%
    }%
    \gplgaddtomacro\gplfronttext{%
      \csname LTb\endcsname%
      \put(3718,3530){\makebox(0,0)[r]{\strut{}$\psi_0$}}%
      \csname LTb\endcsname%
      \put(3718,3310){\makebox(0,0)[r]{\strut{}$\psi_1$}}%
      \csname LTb\endcsname%
      \put(3718,3090){\makebox(0,0)[r]{\strut{}$\psi_2$}}%
      \csname LTb\endcsname%
      \put(3718,2870){\makebox(0,0)[r]{\strut{}$\psi_3$}}%
    }%
    \gplbacktext
    \put(0,0){\includegraphics{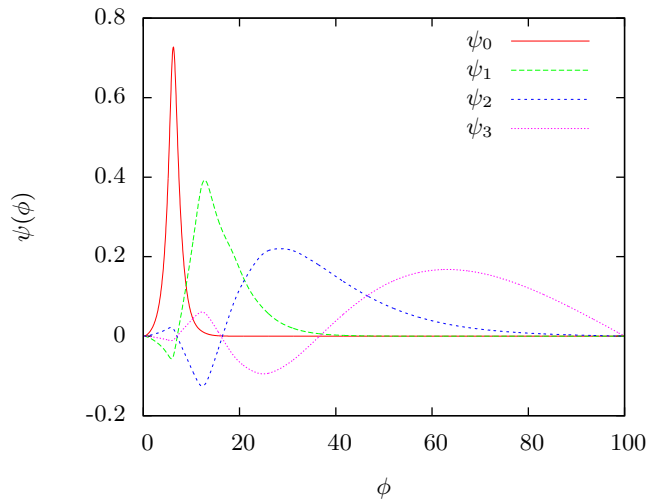}}%
    \gplfronttext
  \end{picture}%
\endgroup
\caption{The wave functions for four lowest energies with $\beta=1$.\label{fig:wavefunctions} Note how the ground state $\psi_0$ has its peak at $\phi=\frac{s_{min}}{\alpha}$, the classical equilibrium distance.}
\end{figure}

We solve the Schr\"{o}dinger equation to obtain the wave function, $\psi$, for the relative motion of the two dipoles in a box of size $\phi\in\left[0:100\right]$, with closed boundary conditions, that is $\psi(0)=\psi(100)=0$. Since the dipole-dipole interaction is 
repulsive at short range the wave function must go to zero as the two particles coincide in space. We may therefore reduce the complexity
of the problem and solve in the region $\phi_1>\phi_2$ only (or vice versa). The full wave function can subsequently be found 
by extended to the opposite domain using continuity and considering the statistics of the particles (even function for bosons and uneven 
function for fermions).
We start out by solving it for the case of $\beta=1$ and $\frac{h}{R}=1$. In figure~\ref{fig:wavefunctions}, the wave functions of the four lowest eigenstates are shown. They all have a negative energy corresponding to bound states of two dipoles on a helix. The solid red curve is the ground state. It has a single maximum, at $\phi\approx2\pi$ corresponding to where the minimum in the two-dipole potential is located. The first excited state is the dashed green curve, it has a single node close to $\phi=10$, between the first two minima of the two-dipole potential. The maximum is close to the second minimum in the potential corresponding to the two dipoles being separated by two windings on the helix.  The short-dashed blue curve is the wavefunction of the second excited state. It has two nodes between the first three minima of the potential, and corresponding maxima at the first two minima, but because it has to be orthogonal to the two lower lying states it has a large part beyond the third minimum. The last curve in dotted purple is the wavefunction of the third excited state. As seen, the curvature at large $\phi$ is negative, whereas the other three had a positive curvature. This is because of the boundary condition at $\phi=100$ where the wavefunction is forced to be zero for our calculations. We will return to this boundary effect below as it will also show up in our calculations of the case with three dipoles on the helix.

\begin{figure}
\centering
\begingroup
  \makeatletter
  \providecommand\color[2][]{%
    \GenericError{(gnuplot) \space\space\space\@spaces}{%
      Package color not loaded in conjunction with
      terminal option `colourtext'%
    }{See the gnuplot documentation for explanation.%
    }{Either use 'blacktext' in gnuplot or load the package
      color.sty in LaTeX.}%
    \renewcommand\color[2][]{}%
  }%
  \providecommand\includegraphics[2][]{%
    \GenericError{(gnuplot) \space\space\space\@spaces}{%
      Package graphicx or graphics not loaded%
    }{See the gnuplot documentation for explanation.%
    }{The gnuplot epslatex terminal needs graphicx.sty or graphics.sty.}%
    \renewcommand\includegraphics[2][]{}%
  }%
  \providecommand\rotatebox[2]{#2}%
  \@ifundefined{ifGPcolor}{%
    \newif\ifGPcolor
    \GPcolortrue
  }{}%
  \@ifundefined{ifGPblacktext}{%
    \newif\ifGPblacktext
    \GPblacktextfalse
  }{}%
  \let\gplgaddtomacro\g@addto@macro
  \gdef\gplbacktext{}%
  \gdef\gplfronttext{}%
  \makeatother
  \ifGPblacktext
    \def\colorrgb#1{}%
    \def\colorgray#1{}%
  \else
    \ifGPcolor
      \def\colorrgb#1{\color[rgb]{#1}}%
      \def\colorgray#1{\color[gray]{#1}}%
      \expandafter\def\csname LTw\endcsname{\color{white}}%
      \expandafter\def\csname LTb\endcsname{\color{black}}%
      \expandafter\def\csname LTa\endcsname{\color{black}}%
      \expandafter\def\csname LT0\endcsname{\color[rgb]{1,0,0}}%
      \expandafter\def\csname LT1\endcsname{\color[rgb]{0,1,0}}%
      \expandafter\def\csname LT2\endcsname{\color[rgb]{0,0,1}}%
      \expandafter\def\csname LT3\endcsname{\color[rgb]{1,0,1}}%
      \expandafter\def\csname LT4\endcsname{\color[rgb]{0,1,1}}%
      \expandafter\def\csname LT5\endcsname{\color[rgb]{1,1,0}}%
      \expandafter\def\csname LT6\endcsname{\color[rgb]{0,0,0}}%
      \expandafter\def\csname LT7\endcsname{\color[rgb]{1,0.3,0}}%
      \expandafter\def\csname LT8\endcsname{\color[rgb]{0.5,0.5,0.5}}%
    \else
      \def\colorrgb#1{\color{black}}%
      \def\colorgray#1{\color[gray]{#1}}%
      \expandafter\def\csname LTw\endcsname{\color{white}}%
      \expandafter\def\csname LTb\endcsname{\color{black}}%
      \expandafter\def\csname LTa\endcsname{\color{black}}%
      \expandafter\def\csname LT0\endcsname{\color{black}}%
      \expandafter\def\csname LT1\endcsname{\color{black}}%
      \expandafter\def\csname LT2\endcsname{\color{black}}%
      \expandafter\def\csname LT3\endcsname{\color{black}}%
      \expandafter\def\csname LT4\endcsname{\color{black}}%
      \expandafter\def\csname LT5\endcsname{\color{black}}%
      \expandafter\def\csname LT6\endcsname{\color{black}}%
      \expandafter\def\csname LT7\endcsname{\color{black}}%
      \expandafter\def\csname LT8\endcsname{\color{black}}%
    \fi
  \fi
  \setlength{\unitlength}{0.0500bp}%
  \begin{picture}(5102.00,3968.00)%
    \gplgaddtomacro\gplbacktext{%
      \csname LTb\endcsname%
      \put(1078,704){\makebox(0,0)[r]{\strut{}-0.2}}%
      \put(1078,1204){\makebox(0,0)[r]{\strut{}-0.15}}%
      \put(1078,1704){\makebox(0,0)[r]{\strut{}-0.1}}%
      \put(1078,2204){\makebox(0,0)[r]{\strut{}-0.05}}%
      \put(1078,2703){\makebox(0,0)[r]{\strut{} 0}}%
      \put(1078,3203){\makebox(0,0)[r]{\strut{} 0.05}}%
      \put(1078,3703){\makebox(0,0)[r]{\strut{} 0.1}}%
      \put(1210,484){\makebox(0,0){\strut{} 0}}%
      \put(1676,484){\makebox(0,0){\strut{} 0.2}}%
      \put(2142,484){\makebox(0,0){\strut{} 0.4}}%
      \put(2608,484){\makebox(0,0){\strut{} 0.6}}%
      \put(3074,484){\makebox(0,0){\strut{} 0.8}}%
      \put(3540,484){\makebox(0,0){\strut{} 1}}%
      \put(4006,484){\makebox(0,0){\strut{} 1.2}}%
      \put(4472,484){\makebox(0,0){\strut{} 1.4}}%
      \put(176,2203){\rotatebox{-270}{\makebox(0,0){\strut{}$E \left[\frac{\hbar^2}{\mu \alpha^2}\right]$}}}%
      \put(2957,154){\makebox(0,0){\strut{}$\beta$}}%
    }%
    \gplgaddtomacro\gplfronttext{%
      \csname LTb\endcsname%
      \put(3718,3530){\makebox(0,0)[r]{\strut{}$E_{0}$}}%
      \csname LTb\endcsname%
      \put(3718,3310){\makebox(0,0)[r]{\strut{}$E_1$}}%
      \csname LTb\endcsname%
      \put(3718,3090){\makebox(0,0)[r]{\strut{}$E_2$}}%
      \csname LTb\endcsname%
      \put(3718,2870){\makebox(0,0)[r]{\strut{}$E_3$}}%
    }%
    \gplbacktext
    \put(0,0){\includegraphics{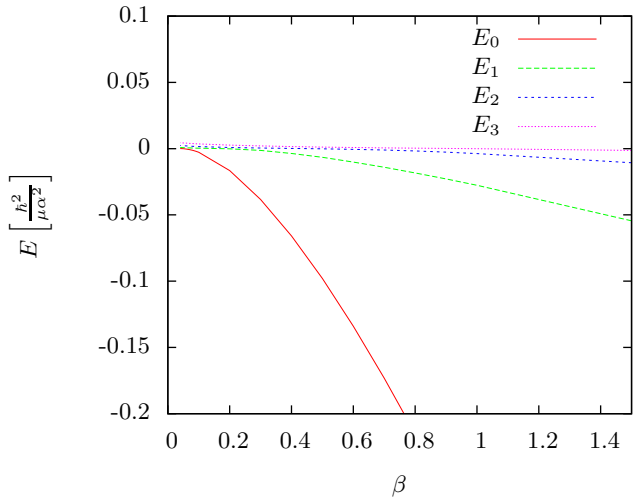}}%
    \gplfronttext
  \end{picture}%
\endgroup
\caption{The lowest four eigenvalues of two dipoles in a helical trap for different potential strengths $\beta$. The energy is in units of $\frac{\hbar^2}{\mu \alpha^2}$.
\label{fig:EofB} }
\end{figure}

In the case of $\beta=1$ shown in figure~\ref{fig:wavefunctions} there are three bound states. In figure~\ref{fig:EofB} the energies of the four lowest states of two dipoles are shown. In the figure one can see that the number of bound states increases with the potential strength as expected. Whether a bound state exists for all positive values of $\beta$ is not possible to answer a priori. Generally in one dimension, if the integral over the potential is negative, that is if $\int{V(\phi)d\phi}\leq 0$ then a bound state exists \cite{landau1936,simon1970,artem2011}. But in our case the potential diverges at short distances and to calculate this integral one would need knowledge of the before mentioned short-range forces that are out of the scope of this paper. The exact potential strength for which individual states become unbound is dependent on the numerical parameters, but the scaling on the number of bound states with $\beta$ is always seen.

\subsection{The Size of the bound states}
\begin{figure}
\centering
\begingroup
  \makeatletter
  \providecommand\color[2][]{%
    \GenericError{(gnuplot) \space\space\space\@spaces}{%
      Package color not loaded in conjunction with
      terminal option `colourtext'%
    }{See the gnuplot documentation for explanation.%
    }{Either use 'blacktext' in gnuplot or load the package
      color.sty in LaTeX.}%
    \renewcommand\color[2][]{}%
  }%
  \providecommand\includegraphics[2][]{%
    \GenericError{(gnuplot) \space\space\space\@spaces}{%
      Package graphicx or graphics not loaded%
    }{See the gnuplot documentation for explanation.%
    }{The gnuplot epslatex terminal needs graphicx.sty or graphics.sty.}%
    \renewcommand\includegraphics[2][]{}%
  }%
  \providecommand\rotatebox[2]{#2}%
  \@ifundefined{ifGPcolor}{%
    \newif\ifGPcolor
    \GPcolortrue
  }{}%
  \@ifundefined{ifGPblacktext}{%
    \newif\ifGPblacktext
    \GPblacktextfalse
  }{}%
  \let\gplgaddtomacro\g@addto@macro
  \gdef\gplbacktext{}%
  \gdef\gplfronttext{}%
  \makeatother
  \ifGPblacktext
    \def\colorrgb#1{}%
    \def\colorgray#1{}%
  \else
    \ifGPcolor
      \def\colorrgb#1{\color[rgb]{#1}}%
      \def\colorgray#1{\color[gray]{#1}}%
      \expandafter\def\csname LTw\endcsname{\color{white}}%
      \expandafter\def\csname LTb\endcsname{\color{black}}%
      \expandafter\def\csname LTa\endcsname{\color{black}}%
      \expandafter\def\csname LT0\endcsname{\color[rgb]{1,0,0}}%
      \expandafter\def\csname LT1\endcsname{\color[rgb]{0,1,0}}%
      \expandafter\def\csname LT2\endcsname{\color[rgb]{0,0,1}}%
      \expandafter\def\csname LT3\endcsname{\color[rgb]{1,0,1}}%
      \expandafter\def\csname LT4\endcsname{\color[rgb]{0,1,1}}%
      \expandafter\def\csname LT5\endcsname{\color[rgb]{1,1,0}}%
      \expandafter\def\csname LT6\endcsname{\color[rgb]{0,0,0}}%
      \expandafter\def\csname LT7\endcsname{\color[rgb]{1,0.3,0}}%
      \expandafter\def\csname LT8\endcsname{\color[rgb]{0.5,0.5,0.5}}%
    \else
      \def\colorrgb#1{\color{black}}%
      \def\colorgray#1{\color[gray]{#1}}%
      \expandafter\def\csname LTw\endcsname{\color{white}}%
      \expandafter\def\csname LTb\endcsname{\color{black}}%
      \expandafter\def\csname LTa\endcsname{\color{black}}%
      \expandafter\def\csname LT0\endcsname{\color{black}}%
      \expandafter\def\csname LT1\endcsname{\color{black}}%
      \expandafter\def\csname LT2\endcsname{\color{black}}%
      \expandafter\def\csname LT3\endcsname{\color{black}}%
      \expandafter\def\csname LT4\endcsname{\color{black}}%
      \expandafter\def\csname LT5\endcsname{\color{black}}%
      \expandafter\def\csname LT6\endcsname{\color{black}}%
      \expandafter\def\csname LT7\endcsname{\color{black}}%
      \expandafter\def\csname LT8\endcsname{\color{black}}%
    \fi
  \fi
    \setlength{\unitlength}{0.0500bp}%
    \ifx\gptboxheight\undefined%
      \newlength{\gptboxheight}%
      \newlength{\gptboxwidth}%
      \newsavebox{\gptboxtext}%
    \fi%
    \setlength{\fboxrule}{0.5pt}%
    \setlength{\fboxsep}{1pt}%
\begin{picture}(4676.00,3400.00)%
    \gplgaddtomacro\gplbacktext{%
      \csname LTb\endcsname%
      \put(682,704){\makebox(0,0)[r]{\strut{}$38$}}%
      \put(682,1312){\makebox(0,0)[r]{\strut{}$40$}}%
      \put(682,1920){\makebox(0,0)[r]{\strut{}$42$}}%
      \put(682,2527){\makebox(0,0)[r]{\strut{}$44$}}%
      \put(682,3135){\makebox(0,0)[r]{\strut{}$46$}}%
      \put(814,484){\makebox(0,0){\strut{}$0$}}%
      \put(1161,484){\makebox(0,0){\strut{}$2$}}%
      \put(1507,484){\makebox(0,0){\strut{}$4$}}%
      \put(1854,484){\makebox(0,0){\strut{}$6$}}%
      \put(2200,484){\makebox(0,0){\strut{}$8$}}%
      \put(2547,484){\makebox(0,0){\strut{}$10$}}%
      \put(2893,484){\makebox(0,0){\strut{}$12$}}%
      \put(3240,484){\makebox(0,0){\strut{}$14$}}%
      \put(3586,484){\makebox(0,0){\strut{}$16$}}%
      \put(3933,484){\makebox(0,0){\strut{}$18$}}%
      \put(4279,484){\makebox(0,0){\strut{}$20$}}%
    }%
    \gplgaddtomacro\gplfronttext{%
      \csname LTb\endcsname%
      \put(176,1919){\rotatebox{-270}{\makebox(0,0){\strut{}$\langle\phi^2\rangle$}}}%
      \put(2546,154){\makebox(0,0){\strut{}$\beta$}}%
    }%
    \gplbacktext
    \put(0,0){\includegraphics{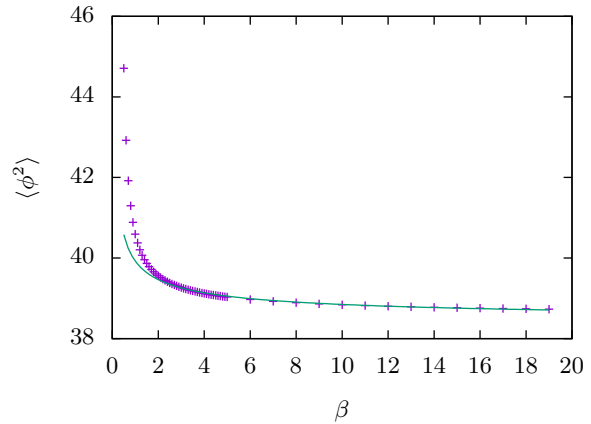}}%
    \gplfronttext
  \end{picture}%
\endgroup
\caption{The expectation value of $\phi^2$ as a function of $\beta$ is shown with plus symbols. The solid line displays a fit of the data to the functional form given in Eq.~\eqref{eq:strongB}. \label{fig:harmonic}}
\end{figure}

In this section we explore how a ground state solution behaves, when the dipole moment changes. Changing the dipole moment corresponds to changing the dimensionless parameter $\beta$.
For a large dipole strength ($\beta$ large), we expect the system to resemble a harmonic oscillator, with the two dipoles oscillating about the distance where the potential is at a minimum. The size of a harmonic oscillator follows the simple relation 
\begin{equation}
<\phi^2>\propto \frac{1}{\omega}\propto\frac{1}{\sqrt{\beta}}
\end{equation}
In this case the harmonic oscillator is not centered around 0 but instead shifted by the equilibrium distance $\phi_0$ so accordingly the value of $<\phi^2>$ should be shifted by $\phi_0^2$, where $\phi_0$ is where $\frac{dV}{d\phi}=0$. So we expect $<\phi^2>$ to scale as
\begin{equation}
<\phi^2>=c_1\frac{1}{\sqrt{\beta}}+c_2\phi_0^2, \label{eq:strongB}
\end{equation}
where $c_1$ and $c_2$ are fitting parameters. We are not interested in their specific value. In figure \ref{fig:harmonic} the size $<\phi^2>$ of the ground state is shown as a function of the parameter $\beta$, while the solid red line is a fit to Eq.~(\ref{eq:strongB}). One sees that in the large $\beta$ limit the size of the ground state behaves as that of a harmonic oscillator.

\begin{figure}
\centering
\begingroup
  \makeatletter
  \providecommand\color[2][]{%
    \GenericError{(gnuplot) \space\space\space\@spaces}{%
      Package color not loaded in conjunction with
      terminal option `colourtext'%
    }{See the gnuplot documentation for explanation.%
    }{Either use 'blacktext' in gnuplot or load the package
      color.sty in LaTeX.}%
    \renewcommand\color[2][]{}%
  }%
  \providecommand\includegraphics[2][]{%
    \GenericError{(gnuplot) \space\space\space\@spaces}{%
      Package graphicx or graphics not loaded%
    }{See the gnuplot documentation for explanation.%
    }{The gnuplot epslatex terminal needs graphicx.sty or graphics.sty.}%
    \renewcommand\includegraphics[2][]{}%
  }%
  \providecommand\rotatebox[2]{#2}%
  \@ifundefined{ifGPcolor}{%
    \newif\ifGPcolor
    \GPcolortrue
  }{}%
  \@ifundefined{ifGPblacktext}{%
    \newif\ifGPblacktext
    \GPblacktextfalse
  }{}%
  \let\gplgaddtomacro\g@addto@macro
  \gdef\gplbacktext{}%
  \gdef\gplfronttext{}%
  \makeatother
  \ifGPblacktext
    \def\colorrgb#1{}%
    \def\colorgray#1{}%
  \else
    \ifGPcolor
      \def\colorrgb#1{\color[rgb]{#1}}%
      \def\colorgray#1{\color[gray]{#1}}%
      \expandafter\def\csname LTw\endcsname{\color{white}}%
      \expandafter\def\csname LTb\endcsname{\color{black}}%
      \expandafter\def\csname LTa\endcsname{\color{black}}%
      \expandafter\def\csname LT0\endcsname{\color[rgb]{1,0,0}}%
      \expandafter\def\csname LT1\endcsname{\color[rgb]{0,1,0}}%
      \expandafter\def\csname LT2\endcsname{\color[rgb]{0,0,1}}%
      \expandafter\def\csname LT3\endcsname{\color[rgb]{1,0,1}}%
      \expandafter\def\csname LT4\endcsname{\color[rgb]{0,1,1}}%
      \expandafter\def\csname LT5\endcsname{\color[rgb]{1,1,0}}%
      \expandafter\def\csname LT6\endcsname{\color[rgb]{0,0,0}}%
      \expandafter\def\csname LT7\endcsname{\color[rgb]{1,0.3,0}}%
      \expandafter\def\csname LT8\endcsname{\color[rgb]{0.5,0.5,0.5}}%
    \else
      \def\colorrgb#1{\color{black}}%
      \def\colorgray#1{\color[gray]{#1}}%
      \expandafter\def\csname LTw\endcsname{\color{white}}%
      \expandafter\def\csname LTb\endcsname{\color{black}}%
      \expandafter\def\csname LTa\endcsname{\color{black}}%
      \expandafter\def\csname LT0\endcsname{\color{black}}%
      \expandafter\def\csname LT1\endcsname{\color{black}}%
      \expandafter\def\csname LT2\endcsname{\color{black}}%
      \expandafter\def\csname LT3\endcsname{\color{black}}%
      \expandafter\def\csname LT4\endcsname{\color{black}}%
      \expandafter\def\csname LT5\endcsname{\color{black}}%
      \expandafter\def\csname LT6\endcsname{\color{black}}%
      \expandafter\def\csname LT7\endcsname{\color{black}}%
      \expandafter\def\csname LT8\endcsname{\color{black}}%
    \fi
  \fi
    \setlength{\unitlength}{0.0500bp}%
    \ifx\gptboxheight\undefined%
      \newlength{\gptboxheight}%
      \newlength{\gptboxwidth}%
      \newsavebox{\gptboxtext}%
    \fi%
    \setlength{\fboxrule}{0.5pt}%
    \setlength{\fboxsep}{1pt}%
\begin{picture}(4676.00,2834.00)%
    \gplgaddtomacro\gplbacktext{%
      \csname LTb\endcsname%
      \put(1078,704){\makebox(0,0)[r]{\strut{}$-0.8$}}%
      \put(1078,891){\makebox(0,0)[r]{\strut{}$-0.75$}}%
      \put(1078,1077){\makebox(0,0)[r]{\strut{}$-0.7$}}%
      \put(1078,1264){\makebox(0,0)[r]{\strut{}$-0.65$}}%
      \put(1078,1450){\makebox(0,0)[r]{\strut{}$-0.6$}}%
      \put(1078,1637){\makebox(0,0)[r]{\strut{}$-0.55$}}%
      \put(1078,1823){\makebox(0,0)[r]{\strut{}$-0.5$}}%
      \put(1078,2010){\makebox(0,0)[r]{\strut{}$-0.45$}}%
      \put(1078,2196){\makebox(0,0)[r]{\strut{}$-0.4$}}%
      \put(1078,2383){\makebox(0,0)[r]{\strut{}$-0.35$}}%
      \put(1078,2569){\makebox(0,0)[r]{\strut{}$-0.3$}}%
      \put(1210,484){\makebox(0,0){\strut{}$-0.02$}}%
      \put(1977,484){\makebox(0,0){\strut{}$-0.015$}}%
      \put(2745,484){\makebox(0,0){\strut{}$-0.01$}}%
      \put(3512,484){\makebox(0,0){\strut{}$-0.005$}}%
      \put(4279,484){\makebox(0,0){\strut{}$0$}}%
    }%
    \gplgaddtomacro\gplfronttext{%
      \csname LTb\endcsname%
      \put(176,1636){\rotatebox{-270}{\makebox(0,0){\strut{}$\left(\langle\phi^2\rangle-\phi_0 ^2\right)\cdot E$ $\left[\frac{\hbar^2}{\mu\alpha^2}\right]$}}}%
      \put(2744,154){\makebox(0,0){\strut{}$E$ $\left[\frac{\hbar^2}{\mu\alpha^2}\right]$}}%
    }%
    \gplbacktext
    \put(0,0){\includegraphics{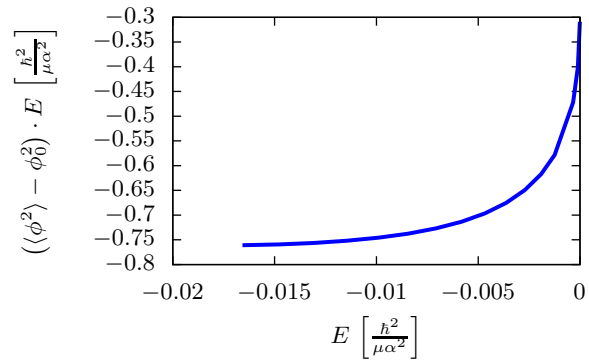}}%
    \gplfronttext
  \end{picture}%
\endgroup
\caption{The expectation value of $\phi^2$ as a function of binding energy.\label{fig:phi2ofE} for the two-body ground state.}
\end{figure}

For small $\beta$ and thus small energies, we expect the system to approach free particles, in the sense that we expect the wave functions to asymptotically approach the form $\Psi(\phi)=e^{-\kappa\phi}$ where $\kappa=\sqrt{2E}$. Using the asymptotic form of the wave function we can calculate the size of the bound states as a function of either the energy $E$, or the strength of the potential $\beta$. Because the wave function for the asymptotic behaviour is not normalized we can calculate $\langle\phi^2\rangle$ using
\begin{equation}
\langle\phi^2\rangle=\frac{\int_{0}^{\infty}\phi^2e^{-2\kappa\phi}d\phi}{\int_{0}^{\infty}e^{-2\kappa\phi}d\phi}=\frac{1}{2\kappa^2}\label{eq:phi2}.
\end{equation} 
So for low energies we expect $\langle\phi^2\rangle$ to scale as $\frac{1}{E}$ according to equation \eqref{eq:phi2}. In figure~\ref{fig:phi2ofE} we show the product of size of the ground state $\langle\phi^2\rangle-\phi_{0}^{2}$ and the ground state energy $E$ as a function of the energy. 

If the scaling follows the $\frac{1}{E}$ we expect this to approach a constant as $E$ decreases. In the figure there is a region 
where it is constant around $E=-0.015$.  However, for very small energies the product increases rapidly. This is because of boundary effects where the size of the ground state approaches the size of the box in which we solve the problem numerically. Alternatively, we see that for the helix sizes we work with the ground state for very small energies becomes so large that the asymptotic behavior is not reached. However, the state does push the boundaries of the geometry as it attempts to extend toward large distances. We have run numerical tests to confirm that once the wave function becomes very broad at low energy, its size scales with the size of the bounding box. 

\section{The Three-body Problem}
We now add a third dipole to the helix. It is of the same type as the other two, that is the same mass and dipole moment. They all interact through the two-dipole interaction in Eq.~\eqref{eq:Vphi}. The position of each of the dipoles is denoted $\phi_1$,$\phi_2$ and $\phi_3$. 
As is typically the case, it is of course very convenient to separate out the center of mass motion of the three dipoles. However, one must keep in mind that we are considering the non-trivial geometry of the helix and it requires some consideration to work this out. The criterion is that all the particles have the same mass, $m$, which is the case we are interested in here. 
The separation can now be done as in other equal-mass three-body systems through the coordinate transformation
\begin{equation}\label{coor}
\begin{split}
x&=\frac{1}{\sqrt{2}}\left(\phi_1-\phi_2\right)\\
y&=\frac{1}{\sqrt{6}}\left(\phi_1+\phi_2\right)-\sqrt{\frac{2}{3}}\phi_3\\
z&=\frac{1}{\sqrt{3}}\left(\phi_1+\phi_2+\phi_3\right).
\end{split}
\end{equation}
With no external potential along the helix, the three body problem now reduces to a two dimensional problem, in the two relative coordinates $x$ and $y$ as the $z$-coordinate only describes a center of mass motion. Note that the three coordinates $x$, $y$, and $z$ are atypical in the sense that they are built from angular variables and thus are intrinsically dimensionless. Without an external potential along the helix, this center of mass motion can be easily worked out. We use open boundary conditions on the helix and thus require the wave functions to vanish on the top and bottom of the helix which implies that the center of mass behaves simply as a particle in a box. 

\begin{figure}
\centering
\begingroup
  \makeatletter
  \providecommand\color[2][]{%
    \GenericError{(gnuplot) \space\space\space\@spaces}{%
      Package color not loaded in conjunction with
      terminal option `colourtext'%
    }{See the gnuplot documentation for explanation.%
    }{Either use 'blacktext' in gnuplot or load the package
      color.sty in LaTeX.}%
    \renewcommand\color[2][]{}%
  }%
  \providecommand\includegraphics[2][]{%
    \GenericError{(gnuplot) \space\space\space\@spaces}{%
      Package graphicx or graphics not loaded%
    }{See the gnuplot documentation for explanation.%
    }{The gnuplot epslatex terminal needs graphicx.sty or graphics.sty.}%
    \renewcommand\includegraphics[2][]{}%
  }%
  \providecommand\rotatebox[2]{#2}%
  \@ifundefined{ifGPcolor}{%
    \newif\ifGPcolor
    \GPcolortrue
  }{}%
  \@ifundefined{ifGPblacktext}{%
    \newif\ifGPblacktext
    \GPblacktextfalse
  }{}%
  \let\gplgaddtomacro\g@addto@macro
  \gdef\gplbacktext{}%
  \gdef\gplfronttext{}%
  \makeatother
  \ifGPblacktext
    \def\colorrgb#1{}%
    \def\colorgray#1{}%
  \else
    \ifGPcolor
      \def\colorrgb#1{\color[rgb]{#1}}%
      \def\colorgray#1{\color[gray]{#1}}%
      \expandafter\def\csname LTw\endcsname{\color{white}}%
      \expandafter\def\csname LTb\endcsname{\color{black}}%
      \expandafter\def\csname LTa\endcsname{\color{black}}%
      \expandafter\def\csname LT0\endcsname{\color[rgb]{1,0,0}}%
      \expandafter\def\csname LT1\endcsname{\color[rgb]{0,1,0}}%
      \expandafter\def\csname LT2\endcsname{\color[rgb]{0,0,1}}%
      \expandafter\def\csname LT3\endcsname{\color[rgb]{1,0,1}}%
      \expandafter\def\csname LT4\endcsname{\color[rgb]{0,1,1}}%
      \expandafter\def\csname LT5\endcsname{\color[rgb]{1,1,0}}%
      \expandafter\def\csname LT6\endcsname{\color[rgb]{0,0,0}}%
      \expandafter\def\csname LT7\endcsname{\color[rgb]{1,0.3,0}}%
      \expandafter\def\csname LT8\endcsname{\color[rgb]{0.5,0.5,0.5}}%
    \else
      \def\colorrgb#1{\color{black}}%
      \def\colorgray#1{\color[gray]{#1}}%
      \expandafter\def\csname LTw\endcsname{\color{white}}%
      \expandafter\def\csname LTb\endcsname{\color{black}}%
      \expandafter\def\csname LTa\endcsname{\color{black}}%
      \expandafter\def\csname LT0\endcsname{\color{black}}%
      \expandafter\def\csname LT1\endcsname{\color{black}}%
      \expandafter\def\csname LT2\endcsname{\color{black}}%
      \expandafter\def\csname LT3\endcsname{\color{black}}%
      \expandafter\def\csname LT4\endcsname{\color{black}}%
      \expandafter\def\csname LT5\endcsname{\color{black}}%
      \expandafter\def\csname LT6\endcsname{\color{black}}%
      \expandafter\def\csname LT7\endcsname{\color{black}}%
      \expandafter\def\csname LT8\endcsname{\color{black}}%
    \fi
  \fi
    \setlength{\unitlength}{0.0500bp}%
    \ifx\gptboxheight\undefined%
      \newlength{\gptboxheight}%
      \newlength{\gptboxwidth}%
      \newsavebox{\gptboxtext}%
    \fi%
    \setlength{\fboxrule}{0.5pt}%
    \setlength{\fboxsep}{1pt}%
\begin{picture}(4676.00,4534.00)%
    \gplgaddtomacro\gplbacktext{%
      \csname LTb\endcsname%
      \put(939,1200){\makebox(0,0){\strut{}$0$}}%
      \put(1502,1061){\makebox(0,0){\strut{}$5$}}%
      \put(2066,922){\makebox(0,0){\strut{}$10$}}%
      \put(2629,783){\makebox(0,0){\strut{}$15$}}%
      \put(2832,864){\makebox(0,0){\strut{}$0$}}%
      \put(3158,1065){\makebox(0,0){\strut{}$5$}}%
      \put(3483,1306){\makebox(0,0){\strut{}$10$}}%
      \put(3809,1586){\makebox(0,0){\strut{}$15$}}%
      \put(878,1850){\makebox(0,0)[r]{\strut{}$0$}}%
      \put(878,2153){\makebox(0,0)[r]{\strut{}$0.2$}}%
      \put(878,2454){\makebox(0,0)[r]{\strut{}$0.4$}}%
      \put(878,2757){\makebox(0,0)[r]{\strut{}$0.6$}}%
      \put(80,2379){\makebox(0,0){\strut{}$\psi(x,y)$}}%
      \put(2338,4115){\makebox(0,0){\strut{}}}%
    }%
    \gplgaddtomacro\gplfronttext{%
      \csname LTb\endcsname%
      \put(1606,892){\makebox(0,0){\strut{}$x$}}%
      \put(3606,1161){\makebox(0,0){\strut{}$y$}}%
    }%
    \gplbacktext
    \put(0,0){\includegraphics{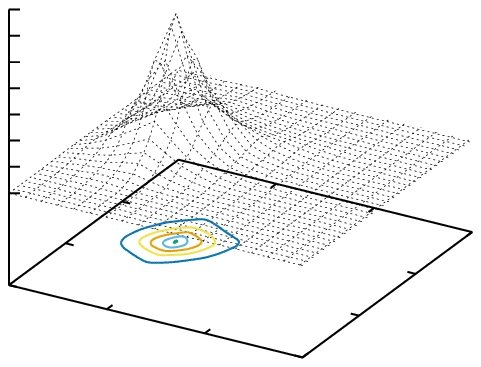}}%
    \gplfronttext
  \end{picture}%
\endgroup
\caption{The ground state wave function of three dipoles on a helix with a potential strength $\beta=1$. The dominant peak corresponds to a configuration where the dipoles are sitting (approximately) one winding apart. Notice that the $x$ and $y$ coordinates are combinations of angles and thus are dimensionless (see Eq.~\eqref{coor}).
\label{fig:3dipolesB1}}
\end{figure}

The two dimensional Schr{\"o}dinger equation that describes the relative motion of the three dipoles should contain five terms, two terms from the kinetic energy of the relative motion between the dipoles, and three potential terms from the pairwise dipole-dipole interaction 
\begin{eqnarray}
& & \tilde H=-\frac{1}{2}\frac{\partial^2}{\partial x^2}-\frac{1}{2}\frac{\partial^2}{\partial y^2}+\beta\tilde V(\sqrt{2}x)\\ \nonumber
&+&\beta\tilde V\left(\sqrt{\frac{3}{2}}y-\frac{1}{\sqrt{2}}x\right)
+\beta\tilde V\left(\sqrt{\frac{3}{2}}y+\frac{1}{\sqrt{2}}x\right)
\end{eqnarray} 
Here $V$ is the two-dipole potential of Eq.(\ref{eq:Vphi}). All terms are written in natural units using $\beta$ as in the case of two dipoles above. The Hamiltonian is also in units of $\mu \alpha^2/\hbar^2$ as in Eq.~\eqref{twoH}. Because the two-dipole potential is repulsive at short range, we use the same trick as for two dipoles to reduce the complexity of the system, we only solve it for $\phi_1>\phi_2>\phi_3$. and then depending on whether the dipoles are bosons or fermions, the full symmetric or antisymmetric wave function can be reconstructed for all values of $\phi_1$,$\phi_2$ and $\phi_3$. The restriction to $\phi_1>\phi_2>\phi_3$ corresponds to a restriction on $x$ and $y$, namely that $y>\frac{1}{\sqrt{3}}x$.
For definiteness, we will be working with a dipolar strength parameter of $\beta=1$ when we plot the wave functions (below we will explore variations with $\beta$ and its effect on the relative distances within the ground state). In addition, we take $h=R$, i.e. the pitch and the radius are the same.
The two-dimensional wave equation is solved on a grid with $x\in[0:160]$ and $y>\frac{1}{\sqrt{3}}x$. 
We observe the same behavior as for two dipoles with regards to the size of the states which also increases with decreasing $\beta$.
From a practical point of view this means that for smaller $\beta$ the size of the grid has to be increased which slow down the numerical solution and reduces accuracy. We therefore restrict ourselves to the case of $\beta=1$ for three dipoles. 

\subsection{Properties of the solutions}
The ground state wave function of the three-body problem is shown in figure~\ref{fig:3dipolesB1} in a three-dimensional level plot that includes a two-dimensional projection onto a contour plot for clarity. The ground state displays a prominent single peak at 
around $x\sim 4.4$ and $y\sim 7.7$. This peak corresponds to a configuration where the three dipoles are placed directly above each other only separated by (approximately) one winding. More precisely, it corresponds to a configuration where $(\phi_1,\phi_2,\phi_3)\sim (4\pi,2\pi,0)$.
We thus see a very nice consistency with respect to the case of two dipoles in which we also find a dominant configuration with the dipoles placed about one winding of the helix apart in order to exploit the attractive interaction in the head-to-tail setup.

\begin{figure*}
\centering
\begingroup
  \makeatletter
  \providecommand\color[2][]{%
    \GenericError{(gnuplot) \space\space\space\@spaces}{%
      Package color not loaded in conjunction with
      terminal option `colourtext'%
    }{See the gnuplot documentation for explanation.%
    }{Either use 'blacktext' in gnuplot or load the package
      color.sty in LaTeX.}%
    \renewcommand\color[2][]{}%
  }%
  \providecommand\includegraphics[2][]{%
    \GenericError{(gnuplot) \space\space\space\@spaces}{%
      Package graphicx or graphics not loaded%
    }{See the gnuplot documentation for explanation.%
    }{The gnuplot epslatex terminal needs graphicx.sty or graphics.sty.}%
    \renewcommand\includegraphics[2][]{}%
  }%
  \providecommand\rotatebox[2]{#2}%
  \@ifundefined{ifGPcolor}{%
    \newif\ifGPcolor
    \GPcolortrue
  }{}%
  \@ifundefined{ifGPblacktext}{%
    \newif\ifGPblacktext
    \GPblacktextfalse
  }{}%
  \let\gplgaddtomacro\g@addto@macro
  \gdef\gplbacktext{}%
  \gdef\gplfronttext{}%
  \makeatother
  \ifGPblacktext
    \def\colorrgb#1{}%
    \def\colorgray#1{}%
  \else
    \ifGPcolor
      \def\colorrgb#1{\color[rgb]{#1}}%
      \def\colorgray#1{\color[gray]{#1}}%
      \expandafter\def\csname LTw\endcsname{\color{white}}%
      \expandafter\def\csname LTb\endcsname{\color{black}}%
      \expandafter\def\csname LTa\endcsname{\color{black}}%
      \expandafter\def\csname LT0\endcsname{\color[rgb]{1,0,0}}%
      \expandafter\def\csname LT1\endcsname{\color[rgb]{0,1,0}}%
      \expandafter\def\csname LT2\endcsname{\color[rgb]{0,0,1}}%
      \expandafter\def\csname LT3\endcsname{\color[rgb]{1,0,1}}%
      \expandafter\def\csname LT4\endcsname{\color[rgb]{0,1,1}}%
      \expandafter\def\csname LT5\endcsname{\color[rgb]{1,1,0}}%
      \expandafter\def\csname LT6\endcsname{\color[rgb]{0,0,0}}%
      \expandafter\def\csname LT7\endcsname{\color[rgb]{1,0.3,0}}%
      \expandafter\def\csname LT8\endcsname{\color[rgb]{0.5,0.5,0.5}}%
    \else
      \def\colorrgb#1{\color{black}}%
      \def\colorgray#1{\color[gray]{#1}}%
      \expandafter\def\csname LTw\endcsname{\color{white}}%
      \expandafter\def\csname LTb\endcsname{\color{black}}%
      \expandafter\def\csname LTa\endcsname{\color{black}}%
      \expandafter\def\csname LT0\endcsname{\color{black}}%
      \expandafter\def\csname LT1\endcsname{\color{black}}%
      \expandafter\def\csname LT2\endcsname{\color{black}}%
      \expandafter\def\csname LT3\endcsname{\color{black}}%
      \expandafter\def\csname LT4\endcsname{\color{black}}%
      \expandafter\def\csname LT5\endcsname{\color{black}}%
      \expandafter\def\csname LT6\endcsname{\color{black}}%
      \expandafter\def\csname LT7\endcsname{\color{black}}%
      \expandafter\def\csname LT8\endcsname{\color{black}}%
    \fi
  \fi
    \setlength{\unitlength}{0.0500bp}%
    \ifx\gptboxheight\undefined%
      \newlength{\gptboxheight}%
      \newlength{\gptboxwidth}%
      \newsavebox{\gptboxtext}%
    \fi%
    \setlength{\fboxrule}{0.5pt}%
    \setlength{\fboxsep}{1pt}%
\begin{picture}(7200.00,5040.00)%
    \gplgaddtomacro\gplbacktext{%
    }%
    \gplgaddtomacro\gplfronttext{%
      \csname LTb\endcsname%
      \put(658,2406){\makebox(0,0){\strut{}$-20$}}%
      \put(1319,2406){\makebox(0,0){\strut{}$-10$}}%
      \put(1980,2406){\makebox(0,0){\strut{}$0$}}%
      \put(2641,2406){\makebox(0,0){\strut{}$10$}}%
      \put(3302,2406){\makebox(0,0){\strut{}$20$}}%
      \put(173,2719){\makebox(0,0)[r]{\strut{}$0$}}%
      \put(173,3380){\makebox(0,0)[r]{\strut{}$10$}}%
      \put(173,4040){\makebox(0,0)[r]{\strut{}$20$}}%
      \put(-157,3528){\rotatebox{-270}{\makebox(0,0){\strut{}$x$}}}%
    }%
    \gplgaddtomacro\gplbacktext{%
    }%
    \gplgaddtomacro\gplfronttext{%
      \csname LTb\endcsname%
    }%
    \gplgaddtomacro\gplbacktext{%
    }%
    \gplgaddtomacro\gplfronttext{%
      \csname LTb\endcsname%
      \put(3898,2406){\makebox(0,0){\strut{}$-20$}}%
      \put(4559,2406){\makebox(0,0){\strut{}$-10$}}%
      \put(5220,2406){\makebox(0,0){\strut{}$0$}}%
      \put(5881,2406){\makebox(0,0){\strut{}$10$}}%
      \put(6542,2406){\makebox(0,0){\strut{}$20$}}%
      \put(7214,2880){\makebox(0,0)[l]{\strut{}$-0.4$}}%
      \put(7214,3204){\makebox(0,0)[l]{\strut{}$-0.2$}}%
      \put(7214,3528){\makebox(0,0)[l]{\strut{}$0$}}%
      \put(7214,3851){\makebox(0,0)[l]{\strut{}$0.2$}}%
      \put(7214,4175){\makebox(0,0)[l]{\strut{}$0.4$}}%
    }%
    \gplgaddtomacro\gplbacktext{%
    }%
    \gplgaddtomacro\gplfronttext{%
      \csname LTb\endcsname%
    }%
    \gplgaddtomacro\gplbacktext{%
    }%
    \gplgaddtomacro\gplfronttext{%
      \csname LTb\endcsname%
      \put(658,264){\makebox(0,0){\strut{}$-20$}}%
      \put(1319,264){\makebox(0,0){\strut{}$-10$}}%
      \put(1980,264){\makebox(0,0){\strut{}$0$}}%
      \put(2641,264){\makebox(0,0){\strut{}$10$}}%
      \put(3302,264){\makebox(0,0){\strut{}$20$}}%
      \put(1980,-66){\makebox(0,0){\strut{}$y$}}%
      \put(173,577){\makebox(0,0)[r]{\strut{}$0$}}%
      \put(173,1238){\makebox(0,0)[r]{\strut{}$10$}}%
      \put(173,1898){\makebox(0,0)[r]{\strut{}$20$}}%
      \put(-157,1386){\rotatebox{-270}{\makebox(0,0){\strut{}$x$}}}%
    }%
    \gplgaddtomacro\gplbacktext{%
    }%
    \gplgaddtomacro\gplfronttext{%
      \csname LTb\endcsname%
    }%
    \gplgaddtomacro\gplbacktext{%
    }%
    \gplgaddtomacro\gplfronttext{%
      \csname LTb\endcsname%
      \put(3898,264){\makebox(0,0){\strut{}$-20$}}%
      \put(4559,264){\makebox(0,0){\strut{}$-10$}}%
      \put(5220,264){\makebox(0,0){\strut{}$0$}}%
      \put(5881,264){\makebox(0,0){\strut{}$10$}}%
      \put(6542,264){\makebox(0,0){\strut{}$20$}}%
      \put(5220,-66){\makebox(0,0){\strut{}$y$}}%
      \put(7214,738){\makebox(0,0)[l]{\strut{}$-0.4$}}%
      \put(7214,1062){\makebox(0,0)[l]{\strut{}$-0.2$}}%
      \put(7214,1386){\makebox(0,0)[l]{\strut{}$0$}}%
      \put(7214,1709){\makebox(0,0)[l]{\strut{}$0.2$}}%
      \put(7214,2033){\makebox(0,0)[l]{\strut{}$0.4$}}%
    }%
    \gplgaddtomacro\gplbacktext{%
    }%
    \gplgaddtomacro\gplfronttext{%
      \csname LTb\endcsname%
    }%
    \gplbacktext
    \put(0,0){\includegraphics{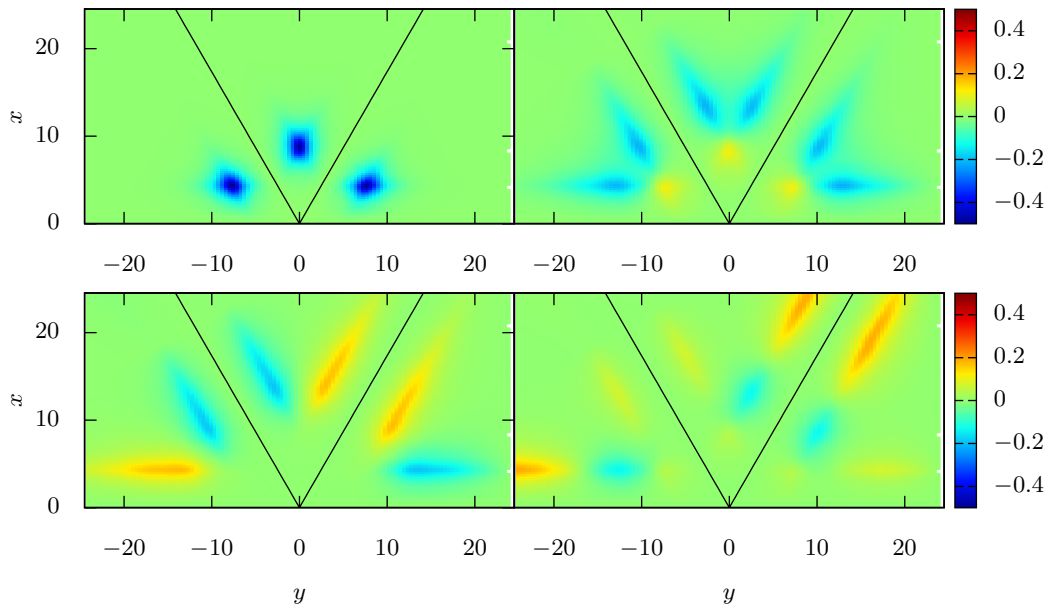}}%
    \gplfronttext
  \end{picture}%
\endgroup
\caption{The wave functions of the four lowest states of three dipoles on a helix shown as contours. The bright (red) regions are positive while the dark (blue) region are negative. Notice that the $x$ axis is vertical, while the $y$ axis is horizontal. The solid (black) lines in panels indicate the lines where $x=\pm\sqrt{3}y$. These are the reflection lines of particle exchange discussed in the text. Here the particles are assumed to obey bosonic statistics (see the text). The ground state (top left), the first excited state (top right), the second excited state (bottom left), and the third excited state (bottom right) are shown. Note again that $x$ and $y$ are dimensionless as in figure~\ref{fig:3dipolesB1}.}
\label{fig:5lowest} 
\end{figure*}

We now consider the low-energy spectrum of three dipoles on a helix by representing the wave functions in contour plots as functions of $x$ and $y$ coordinates. This is shown in figure~\ref{fig:5lowest}. As discussed above, the wave equation has been solved only for $\sqrt{3}y>x$. Solid (black) lines indicate where $x=\pm\sqrt{3}y$ in all the panels in figure~\ref{fig:5lowest}. To extend the results to the full $(x,y)$-coordinate space, one needs to reflect the wave functions across the solid lines in the panels. 
Here we are assuming that the dipoles are identical bosons. This implies that the wave function must be symmetric in the exchange of any pair of particles, and thus we must reflect across the solid lines and keep a positive sign.
In the case of fermionic dipoles, the solution would be the same except for a sign change across the diagonal.

The panels in figure~\ref{fig:5lowest} show the first four states in the spectrum. In the upper left corner, we have the ground state which is identical to the state shown in a different manner in figure~\ref{fig:3dipolesB1}. It is a state with three head-to-tail dipoles, here extended across the solid lines so it is really three copies of the central region (within the wedge traced by the solid lines containing the $y=0$ axis). The first excited state in the top right corner has a much more detailed structure. It still contains a trace of the head-to-tail on successive windings structure of the ground state but now a lot of amplitude is move to larger values of $x$ in the central region around $y=0$ (within the central wedge). This corresponds to configurations where the one of the outer dipoles from the head-to-tail configurations is now pushed one winding away from the two others (this can be done in two ways so we have $y\to -y$ symmetry here). Notice also the change in sign between the two regions with non-zero amplitudes. This is of course a result of the fact that higher excited states have additional nodes in the wave functions. We may relate this very directly to the physics we saw in the case of two dipoles in figure~~\ref{fig:wavefunctions}. There we see that the first excited two-body state (long dashed (green) line) has a small bit of amplitude around angles of $\phi\sim 2\pi$ but that most of its weight is around $\phi\sim 4\pi$ and thus the two-body state has the dipoles sitting about two windings apart mainly. This is clearly also reflected in the contour plot of the first excited state for three dipoles seen in the top right panel of figure~\ref{fig:5lowest}.

The second excited state of three dipoles is shown in the lower left panel in figure~\ref{fig:5lowest} and has a structure that can now be simply understood given the two lower states. Its amplitude is dominated by pairs of dipoles which are now about two winding apart. In comparison to the first excited state it has zero amplitude of the simple head-to-tail configuration seen in the ground state. It is a bit harder to compare this to the two-body case as the second excited state for two dipoles has them sitting really far apart (at the level of $\phi\sim 5\pi$, see figure~\ref{fig:wavefunctions}), while the second excited state for three dipoles resembles more the physics in the first excited state in the two-body case only without the configurations where two of the dipoles are sitting one winding apart. The third excited state seen in the lower right panel of figure~\ref{fig:5lowest} tells a similar story except that now the dipoles move even further apart as the amplitude is seen to move to larger values of $x$ and $y$. We also notice that a small bit of amplitude comes back to the ground state configuration in the region near the origin $(x,y)=(0,0)$. However, the third excited state is not fully converged as one can see by the lack of symmetry (up to a sign) for $y\to -y$. It is the same sort of boundary effect that can be seen in figure~\ref{fig:wavefunctions} for the third excited state of the two-body system. In spite of this numerical issue, the tendency of higher excited states should now be clear. In higher excited states the dipoles are pushed further and further away from each other and is consistent with the picture that we have from the two-body case. A nice feature is that the ground state configuration with three dipoles head-to-tail on three successive windings does indeed seems to make an appearance in higher excited states also so we do see that the first minimum in the two-body potential 
in figure~\ref{fig:potential} plays a very dominant role in this geometry.

In order to further elucidate the configuration of the three dipoles of the helix we can consider the relative
distances in angle between each pair of dipoles within the ground state. These distances are defined as
$\phi_{ij}=\phi_i-\phi_j$ and we take the expectation value of this operator in the ground state. The 
results are shown in table~\ref{tab:Slist} for different values of $\beta$ with $h=R$. 
The distances are calculated in units of $2\pi$. 
Because of the chosen ordering they are all positive, and $\phi_{13}=\phi_{12}+\phi_{23}$. 
As seen in the table, for larger values of $\beta$ the ground state has a clear interpretation as a 'chain' of three dipoles separated by one winding. This is seen to set in already for $\beta=1$ and is accurate at the level of two decimal places already for $\beta=2$. For smaller values of $\beta$ we expect that kinetic terms will be more important and the dipoles would like to delocalize. In the table this is seen for a value of $\beta=0.25$ where the expectation values of the distances are no longer
close to $2\pi$, i.e. the dipoles tend to be sitting further than one winding apart on average. In this regime of weak 
dipolar interactions the dipoles will tend to spread out to minimize kinetic energy while at the same time being able to take
advantage of the attraction from several of the pockets seen in figure~\ref{fig:potential}.

\begin{table}[h!]
\caption{Relative angular distances in the ground state for three dipoles 
with $h=R$ in units of $2\pi$ for different values of $\beta$.\label{tab:Slist}}
\begin{ruledtabular}
\begin{tabular}{c| c c c c c}
$\beta$	&$\langle \phi_{12}\rangle/2\pi$		&$\langle \phi_{23}\rangle/2\pi$	&$\langle \phi_{13}\rangle/2\pi$\\
\hline
$\beta=0.25$		&$1.50$	&$1.44$	&$2.94$	\\ 
$\beta=1$			&$1.01$	&$1.01$	&$2.03$	\\
$\beta=2$			&$1.00$	&$1.00$	&$2.00$	 \\
\end{tabular}
 \end{ruledtabular}
\end{table}
 
\section{Discussion And Outlook}
In the present paper we have considered the physics of dipolar particles that are confined to move on a one-dimensional helix.
We first look at the dipole-dipole interaction on a helix, and how the interplay between the long range interaction and the peculiar geometry of the helix leads to a two particle potential with several minima of decreasing depth. These minima correspond to the attractive head-to-tail configuration and the decreasing depth of successive minima is a result of the dipoles being an increasing number of winding of the helix apart. Our main question concerns the formation of bound state in this non-trivial system. In particular, what the formation criteria for two-body bound states are and what type of bound states form with more dipoles. As our main focus we use the case of three dipoles.

In the strong interaction limit both two- and three-body bound states correspond to the dipoles being an integer number of windings apart in the quantum ground state of the system. However, for weaker interactions the bound states increase in size, and pairs of dipoles can no longer be said to be a certain number of windings apart as they become effectively delocalized across several windings of the helix. For three dipoles we show that for moderate dipole strengths they form a short yet well-defined chain of three dipoles sitting immediately underneath each other. 

This affinity for chain formation was discussed previously in the limit of very strong interactions where classical crystal formation on the helix is expected \cite{pedersen2014}. 
There it was shown how the dipoles would form chains up the helix where each dipole was approximately one winding away from its neighbors on either side. In one dimension such long range orders are not possible in the quantum regime, and one would instead expect the formation of a Luttinger liquid in such a system as has been discussed in Ref.~\cite{law2008}. This should be more pronounced for smaller values of $\beta$ where kinetic terms are sizable and the particles will tend to delocalize, i.e. not merely stay approximately fixed in the minima provided by the attractive head-to-tail configuration. The formation of 
chains in related geometries in both two- \cite{wang06,arms-dip12} and one-dimensional \cite{klawunn2010,artem-dip2013} setups, and this chain formation is expected to persist and be in the many-body case also in the quantum regime \cite{wang06,barbara2011}. The strongly interacting regime can be explored by using harmonic approximations \cite{arms11} to the full dipolar interaction for instance in the study of the thermodynamic properties of dipolar chains \cite{arms-dip13}. As we discussed above, a harmonic behavior is also seen in the helical geometry for large $\beta$ which is merely a reflection of the fact that the dipolar potential allows such an approximation for large dipole moments in any geometry where the head-to-tail configuration is possible. In future studies it would be interesting to extend the system to slightly longer chains and study the thermodynamics for instance using the harmonic approximation for strong interactions.

The authors acknowledge
support from the Danish Council for Independent Research and 
the DFF Sapere Aude program.

\end{document}